\documentclass[onecolumn,showpacs,aps,floatfix,pra,superscriptaddress,nofootinbib]{revtex4}
\usepackage{color}
\usepackage{graphicx}
\usepackage{bm}
\usepackage{amsmath}
\begin{document}
\title{Scattering by a reflectionless modified P\"{o}schl-Teller potential:
Bohmian trajectories and arrival times}
\author{S. V. Mousavi}
\email{vmousavi@qom.ac.ir} \affiliation{Department of Physics, The
University of Qom, P. O. Box 37165, Qom, Iran} \affiliation{School
of Physics, Institute for Research in Fundamental Sciences (IPM),
P.O.Box 19395-5531, Tehran, Iran }

%%%%%%%%%%%%%%%%%%%%%%%%%%%%%%%%%%%%%%%%%%%%%%%%%%%%%%%%%%%%%%%%%%%

%
\begin{abstract}

A nonreflecting wavepacket is constructed by the superposition of
reflectionless eigenstates of sech2 potential. Free propagation and
propagation in the presence of the above potential of such a
wavepacket is considered using the concept of arrival time.
Comparison is made with the free evolving Gaussian wavepacket. Mean
arrival time at a detector behind the well is given as a function of
mass for separate cases. A selection of Bohmian trajectories in the
interacting case are computed and compared to the trajectories of a
free particle guided by a Gaussian packet.

\end{abstract}
 \maketitle

%%%%%%%%%%%%%%%%%%%%%%%%%%%%%%%%%%%%%

%== Section ==== Section ==== Section ==== Section ==== Section ==== Section ==== Section ==== Section ==== Section ==== Section

\section{Introduction}

Scattering of incident waves by inhomogeneities is a common physical
feature in optics and quantum mechanics.
The trigonometric P\"{o}schl-Teller (PT) potential was suggested by
P\"{o}schl and Teller \cite{PT-ZP-1933} to describe diatomic
molecular vibration.
PT-type potentials have been noticed by authors from different areas
of physics ranging from the molecular physics to the nuclear one.
Using different approximation methods, analytical solutions of the
Schr\"{o}dinger equation and energy eigenvalues for diatomic
molecules have been found for the trigonometric PT molecular
potential \cite{Fa-CJP-2012}.
Approximate bound-state solutions of the Klein-Gordon
\cite{PLA2008-PS2009} and Dirac \cite{JPA2008-EPJA2010} equations
with such potentials have been investigated.
The Sech-squared potential (or Modified PT potential) has attracted
too too much attention, for instance it appears in the context of
neutron-matter \cite{GeCa-PRC-2010} and spherical quantum dot
\cite{HaKaTe-SM-2013}.
There are obstacles, called reflectionless potentials,
which are completely transparent for the incident wave. This type of
potentials have been extensively studied analytically and
numerically in the literature.
%
%The study of this astonishing topic is practically
%important.
%
Using the method of summation over eigenstates, the exact propagator
for a general reflectionless potential has been computed
\cite{Cr-JPA-1983}.
Scattering of a Gaussian wavepacket by a reflectionless sech2
potential has been numerically studied and possible applications of
such potentials has been argued \cite{KiKu-AJP-1998}.
It has been shown that in comparison to the case of free evolution,
the wavepacket accelerates in the well and is narrower.
Authors arrived at this result by
plotting the probability density versus position in a time that the
whole wavepacket has left the well.
In this paper we employ the concept of arrival time \cite{MuLe-PR-2000}
for quantitative study of this problem.
In the case of reflectionless sech2 potential non-reflecting
wavepackets have been constructed by superposing energy eigenstates
and propagation of them has been studied in the presence of the
above potential \cite{Le-AJP-2007}.
Similar results have been reproduced by the method of supersymmetric
quantum mechanics \cite{CoLe-EJP-2008}.
Traversal time through the reflectionless sech2 potential has been
studied using the method of a quantum clock \cite{Pa-PLA-2011}.
Only, recently, a true reflectionless potential was realized
experimentally in photonic lattices using the concept of arrays of
evanescently couple waveguides \cite{Szetal-PRL-2011}.

It seems to be interesting to consider the problem of a wavepacket
scattering by the reflectionless Sech2 potential using a
trajectory-based theory, since by attributing a well-defined
trajectory to a particle, its arrival time at a detector location is
a well-defined and predictable quantity. It has been shown within
the Bohmian mechanics \cite{BoHi-book-1993} that the arrival time
distribution is given by the probability current density
\cite{Le-PLA-1993}. But, there is no consensus about this quantity
within the standard formulation of the quantum mechanics
\cite{Fi-PRA-199}.

%To the best of our knowledge, no one has yet considered scattering of
%wavepackets by a reflectionless sech2 potential from the Bohmian
%point of view:
%
%an alternative interpretation that gives physical account of individual phenomena
%and leads to the same statistical results as does quantum mechanics \cite{BoHi-book-1993}.
%
To this end, we at first follow Lekner \cite{Le-AJP-2007} to
construct a non-reflecting wavepacket. Then, propagation of such a
constructed wavepacket is considered and arrival time at a detector
location behind the well is computed. Comparison with a freely
evolving Gaussian wavepacket is made. Finally, a selection of
Bohmian paths is plotted.

%== Section ==== Section ==== Section ==== Section ==== Section ==== Section ==== Section ==== Section ==== Section ==== Section

\subsection{Constructing a nonreflecting wavepacket}

Exact solution of the Schr\"{o}dinger equation for any positive
energy $ E_k = \frac{\hbar^2 k^2}{2m}$ is given by
\begin{eqnarray} \label{eq: sch_exact_sol}
\psi_k(x) &=&  A(k) \frac{ i k - a \tanh(ax) }{ik+a} e^{ikx} ~,
\end{eqnarray}
in the presence of the sech2 potential
\begin{eqnarray} \label{eq: sech2pot}
V(x) &=& -\frac{\hbar^2  a^2}{2 m} \frac{\nu(\nu+1)}{\cosh^2(a x)} ~,
\end{eqnarray}
with $ \nu = 1 $. One easily sees that in the limit $ x \rightarrow
-\infty $, the above solution reduces to $ A(k) e^{ikx}$. This shows
the nonreflectivity of the potential; with an incident wave from the
left there is no reflected wave \cite{Gr-book-1995}.
%It is noteable that the eigenfunction (3) of ref. \cite{CoLe-EJP-2008}
%does not have this property even though it solves the Schr\"{o}dinger equation.

By superposing the reflectionless eigenstates (\ref{eq:
sch_exact_sol}), one obtains nonreflecting wavepacket
\begin{eqnarray} \label{eq: nonref_packet_integral}
\Psi_{\text{i,nr}}(x, t) &=& \frac{1}{ \sqrt{2\pi} }
\int_{-\infty}^{\infty} dk~ \psi_k(x) e^{ -i E_k t / \hbar} ~,
\end{eqnarray}
where sub-indices "i" and "nr" stand respectively for "interacting"
and "nonreflecting". Noting the factor $(ik+a)^{-1}$ in the relation
of $ \psi_k(x) $, a simple choice for $A(k)$ is \cite{Le-AJP-2007}
\begin{eqnarray} \label{eq: A(k)}
A(k) &=& (ik+a) \phi_0(k) ~,
\end{eqnarray}
where,
\begin{eqnarray} \label{eq: Fourier_Gaussian}
\phi_0(k) &=& \left( \frac{2}{\pi} \sigma_0^2 \right)^{1/4}
e^{-\sigma_0^2(k-k_0)^2} e^{ - i (k-k_0) x_c } ~,
\end{eqnarray}
is the Fourier transform of the initial Gaussian wavepacket
\begin{eqnarray} \label{eq: initial_Gaussian}
\Psi_G(x, 0) &=& \frac{1}{(2\pi \sigma_0^2)^{1/4}} \exp \left[
-\frac{(x-x_c)^2}{4\sigma_0^2} + i k_0 x \right]~,
\end{eqnarray}
where subindex "G" stands for "Gaussian". Here, $k_0$ is the kick
momentum; and $x_c$ and $\sigma_0$ are respectively the center and
rms width of the packet. Time-evolved Gaussian wavepacket in free
space is given by,
\begin{eqnarray} \label{eq: free_Gaussian}
\Psi_{\text{f,G}}(x, t) &=& \frac{1}{(2\pi s_t^2)^{1/4}} \exp \left[
-\frac{(x-x_c-ut)^2}{4 s_t \sigma_0} + i k_0 \left( x - \frac{ut}{2}
\right) \right]~, ~~~~~ s_t = \sigma_0 \left( 1 + i \frac{\hbar
t}{2m \sigma_0^2} \right)~,
\end{eqnarray}
where $s_t = \sigma_0(1+i\hbar t/2m\sigma_0^2)$ and $ u = \hbar k_0
/ m $ is the group velocity of the packet. In this equation subindex
"f" stands for "free".

By the choice (\ref{eq: A(k)}) for $ A(k) $, reflectionless wavepacket reads
\begin{eqnarray} \label{eq: nonref_packet}
\Psi_{\text{i,nr}}(x, t) &=& N \left[  i k_0 - \frac{ x-x_c-ut }{ 2
\sigma_0 s_t } - a \tanh(a x) \right ] \Psi_{\text{f,G}}(x, t)  ~,
\end{eqnarray}
where $N$, which has the length dimension, shows the normalization constant.
Free evolution of the initial nonreflecting wavepacket $
\Psi_{\text{nr}}(x) = \Psi_{\text{i,nr}}(x, 0)$ is given by the
integral relation
\begin{eqnarray} \label{eq: free_nonref_packet}
\Psi_{\text{f,nr}}(x, t) &=& \int_{-\infty}^{\infty} dx^{\prime}~
G_{\text{f}}(x, t; x^{\prime}, 0) \Psi_{\text{nr}}(x^{\prime}) ~,
\end{eqnarray}
where,
\begin{eqnarray*} \label{eq: free_propagator}
G_{\text{f}}(x, t; x^{\prime}, 0) &=& \sqrt{ \frac{m}{2\pi i \hbar
t} } \exp \left\{\frac{im}{2\hbar t}(x-x^{\prime})^2 \right\}~,
\end{eqnarray*}
represents the propagator of free particle. From eqs. (\ref{eq:
nonref_packet}) and (\ref{eq: free_nonref_packet}) one obtains,
\begin{eqnarray} \label{eq: free_nonref}
\Psi_{\text{f,nr}}(x, t) &=& N \bigg\{ i k_0 - \frac{x-x_c-u
t}{2\sigma_0 s_t} -a \sqrt{ \frac{m}{2\pi i \hbar t}
\frac{s_t}{\sigma_0} }  ~ g(x, t) \bigg\} \Psi_{\text{f,G}}(x, t)~,
\end{eqnarray}
where,
\begin{eqnarray*}
g(x, t) &=& \int_{-\infty}^{\infty} dx^{\prime} ~ \tanh (a
x^{\prime}) \exp \left \{ -\frac{im}{2\hbar t} \frac{s_t}{\sigma_0}
\left[ x^{\prime} - \frac{\sigma_0}{s_t} \left( x-ut+\frac{i\hbar
t}{2m\sigma_0^2} x_c \right)  \right]^2 \right \} ~.
\end{eqnarray*}

Just for completeness we mention that the evolution of the
Gaussian wavepacket (\ref{eq: initial_Gaussian}) in the presence of
the nonreflecting potential (\ref{eq: sech2pot}) with $\nu=1$ which
has been numerically studied in \cite{KiKu-AJP-1998}, is given by
\begin{eqnarray} \label{eq: interacting_Gaussian_packet}
\Psi_{\text{i,G}}(x, t) &=& \int_{-\infty}^{\infty} dx^{\prime}~
G_{\text{i}}(x, t; x^{\prime}, 0) \Psi_G(x^{\prime}, 0) ~,
\end{eqnarray}
where,
\begin{eqnarray*} \label{eq: interacting_propagator}
G_{\text{i}}(x, t; x^{\prime}, 0) &=& G_{\text{f}}(x, t; x^{\prime},
0) + \frac{ \exp \left[\frac{i\hbar t}{2 m} a^2 \right] }{2 \cosh(a
x) \cosh(a x^{\prime})}
\nonumber \\
& \times & \left\{ {\text{erf}} \left( a\sqrt{ \frac{i \hbar t}{2
m}} - (x-x^{\prime}) \sqrt{ \frac{m}{2i\hbar t} } \right) +
{\text{erf}} \left( a\sqrt{ \frac{i \hbar t}{2 m}} + (x-x^{\prime})
\sqrt{ \frac{m}{2i\hbar t} } \right) \right \} ~,
\end{eqnarray*}
represents the propagator of the nonreflecting potential for $\nu=1$
\cite{Cr-JPA-1983, GrSt-book-1998}.
%

%== Section ==== Section ==== Section ==== Section ==== Section ==== Section ==== Section ==== Section ==== Section ==== Section

\section{Basic equations} \label{sec: basic_eqs}

Using eq. (\ref{eq: nonref_packet}) probability density and
probability current density becomes
\begin{eqnarray} \label{eq: nonref_rho}
\rho_{\text{i,nr}}(x, t) &=& |N|^2 \left[  \left( k_0 + \frac{\hbar
t (x-x_c-ut)}{4m \sigma_0^2 \sigma_t^2} \right)^2 + \left( a \tanh(a
x) + \frac{ x-x_c-ut }{2 \sigma_t^2}  \right)^2 \right]
\nonumber \\
& \times & \frac{ 1 }{ \sqrt{2 \pi} \sigma_t } \exp \left[ -
\frac{(x-x_c-ut)^2}{2 \sigma_t^2} \right] ~,
\end{eqnarray}
and
\begin{eqnarray} \label{eq: nonref_cur}
j_{\text{i,nr}}(x, t) &=& |N|^2 \frac{ m \sigma_0 \hbar
\sqrt{\frac{2}{\pi }} \bigg( f_3(x) ~ m^3 + f_2(x,t) ~ m^2 +
f_1(x,t) ~ m + f_0(x, t) \bigg) }{\left(2 m \sigma_0 \sigma_t
\right)^5} \exp \left[ - \frac{(x-x_c-ut)^2}{2 \sigma_t^2} \right]~,
\end{eqnarray}
where $ \sigma_t = |s_t| = \sigma_0 \sqrt{ 1 + \frac{\hbar^2
t^2}{4m^2 \sigma_0^4} } $ and
\begin{eqnarray*}
f_3(x) &=& 4 \bigg[ 4 k_0^3 \sigma_0^8 + k_0 \sigma_0^4 (2
\sigma_0^2 + 4 a^2 \sigma_0^4 + (x - x_c)^2) + 4 a k_0 \sigma_0^6 (x
- x_c) \tanh(a x)\bigg]~,
\\
f_2(x, t) &=& 4 \hbar t \bigg[ a^2 \sigma_0^4 (x - x_c) + k_0^2
\sigma_0^4 (x - x_c) -
 a \sigma_0^2 (\sigma_0^2 + 4 k_0^2 \sigma_0^4 - (x - x_c)^2) \tanh(a x) + (x-x_c)^3  \bigg]~,
\\
f_1(x, t) &=& \hbar^2 t^2 \bigg[ 2 k_0 \sigma_0^2 (1 + 2 a^2
\sigma_0^2) -
 4 a k_0 \sigma_0^2 (x - x_c) \tanh(a x) \bigg] ~,
\\
f_0(x, t) &=& a \hbar^3 t^3 \bigg[ a (x - x_c) - \tanh(a x) \bigg]~.
\end{eqnarray*}

According to the Ehrenfest theorem, expectation value of momentum
remains unchanged in the free space. Thus, in the free space one has
\begin{eqnarray} \label{eq: free_<x>}
\langle x \rangle(t) &=&  \langle x \rangle(0) + \frac{ \langle p \rangle(0) }{ m } t ~.
\end{eqnarray}
For the freely evolving Gaussian wavepacket this equation reads,
\begin{eqnarray} \label{eq: free_<x>_G}
\langle x \rangle_{\text{f,G}}(t) &=&  x_c + u t ~.
\end{eqnarray}

In the Bohmian view of the world, the state of a system is described
by its wave function together with its actual configuration defined
by the actual position of its particle. In the minimal approach to
the theory \cite{DuGoZa-JSP-1992}, particle equation of motion,
known as guidance equation, is
\begin{eqnarray} \label{eq: v_bohm}
v(x, t) &=& \frac{j(x, t)}{\rho(x, t)} = \frac{\hbar}{m} \Im \frac{
\nabla \psi}{ \psi}~.
\end{eqnarray}
Using this equation and eqs. (\ref{eq: nonref_rho}) and (\ref{eq: nonref_cur}), one obtains
\begin{eqnarray} \label{eq: vb_initial_nonref}
v_{\text{nr}}(x, 0) &=& u~\frac{1}{16 \sigma_0^8 k_0} \frac{ f_3(x)
} { k_0^2 + \left( a \tanh(a x) + \frac{ x-x_c }{2 \sigma_0^2}
\right)^2 } ~,
\end{eqnarray}
for the initial Bohmian velocity of a particle which is guided by
the nonreflecting wave packet (\ref{eq: nonref_packet}). It is clear
from eq. (\ref{eq: vb_initial_nonref}) that initial velocity depends
on the position in the nonreflecting wavepacket case, while it is
constant value $u$ for the Gaussian packet.
By integrating guidance equation (\ref{eq: v_bohm}) over time one
obtains
\begin{eqnarray} \label{eq: bohm_tarj_G}
x_{\text{f,G}}(t) &=& x_c + u t + \frac{ \sigma_t }{ \sigma_0 }(x_0 - x_c)~,
\end{eqnarray}
for the trajectories of a free particle which is guided by the
Gaussian packet (\ref{eq: free_Gaussian}). Here, $x_0$ is the
initial position of the particle.

Within the Bohmian mechanics, the arrival time distribution of
particles at detector location $x_d$ is given by \cite{Le-PLA-1993}
\begin{eqnarray} \label{eq: arr_dis}
\Pi(x_d, t) &=& \frac{ |j(x_d, t)| }{ \int_0^{\infty} dt~|j(x_d, t)| } ~,
\end{eqnarray}
and thus the mean arrival time reads
\begin{eqnarray} \label{eq: mean_arr}
\tau(x_d) &=& \int_0^{\infty} dt~ t ~\Pi(x_d, t) ~.
\end{eqnarray}
%

%== Section ==== Section ==== Section ==== Section ==== Section ==== Section ==== Section ==== Section ==== Section ==== Section

\section{ Scaling the unit of length }

In order to do numerical calculations, we apply a scaling
transformation which eliminates physical constant $\hbar$ and mass
$m$ in the Schr\"{o}dinger equation for free particle and for a
particle interacting with sech2 potential (\ref{eq: sech2pot}).
Wavefunction $ \Psi(x, t) $ which describes a particle of mass $ m $
in the potential (\ref{eq: sech2pot}) with $ \nu = 1 $, satisfies
the Schr\"{o}dinger equation
\begin{eqnarray} \label{eq: Sch eq}
i\hbar \frac{\partial}{\partial t} \Psi(x, t) &=&  \left( -\frac{\hbar^2}{2m} \frac{\partial^2}{\partial x^2}  -
\frac{\hbar^2 a^2}{m} {\text{sech}}^2( a x)\right) \Psi(x, t)~.
\end{eqnarray}
By space transformation $ x \rightarrow \bar{x} = \sqrt{ \frac{m}{\hbar} } x $,
which is a scaling of unit length \cite{Th-book-2000},
and defining new function
\begin{eqnarray} \label{eq: new wf}
\bar{\Psi}(\bar{x}, t) &=& \Psi(\sqrt{ \hbar / m }~\bar{x}, t) ~,
\end{eqnarray}
Schr\"{o}dinger equation (\ref{eq: Sch eq}) reads
\begin{eqnarray} \label{eq: modifed Sch eq}
i \frac{\partial}{\partial t} \bar{\Psi}(\bar{x}, t) &=&  \left( -\frac{1}{2} \frac{\partial^2}{\partial \bar{x}^2}  -
\bar{a}^2 {\text{sech}}^2(\bar{a} \bar{x})\right) \bar{\Psi}(\bar{x}, t)~,
\end{eqnarray}
where $ \bar{a} = \sqrt{ \frac{\hbar}{m} } a $. Thus, transformed
momentum operator is $ \bar{p} = - i \partial / \partial \bar{x}$.
Eq. (\ref{eq: modifed Sch eq}) says that the above scale
transformation corresponds to put $\hbar = m = 1$ in the equations
of section \ref{sec: basic_eqs}.
%During this process every quantity of length unit is replaced by
%$ \sqrt{\hbar/m } $ times that quantity, and
%every quantity of inverse length unit transforms to $ \sqrt{m/\hbar } $ times the quantity.

Continuity equation becomes
\begin{eqnarray} \label{eq: transformed-continuity}
\frac{\partial \bar{\rho}(\bar{x}, t)}{\partial t} + \frac{\partial \bar{j}(\bar{x}, t)}{\partial \bar{x}} &=& 0  ~,
\end{eqnarray}
where,
\begin{eqnarray} \label{eq: transfomed rho&j}
\bar{\rho}(\bar{x}, t) &=& | \bar{\Psi}(\bar{x}, t) |^2 ~,\nonumber \\
\bar{j}(\bar{x}, t) &=& \Im \left\{ \bar{\Psi}^*(\bar{x}, t) \frac{\partial}{\partial \bar{x}} \bar{\Psi}(\bar{x}, t) \right\} ~,
\end{eqnarray}
and thus, guidance equation in this new coordinate reads
\begin{eqnarray} \label{eq: new normalization}
\bar{v}(\bar{x}, t) &=& \frac{ \bar{j}(\bar{x}, t) }{ \bar{\rho}(\bar{x}, t) }  ~.
\end{eqnarray}
%

%The normalization condition is as usual,
%
%\begin{eqnarray} \label{eq: new normalization}
%\int_{-\infty}^{\infty} d\bar{x} ~ \bar{\rho}(\bar{x}, t) &=& 1~.
%\end{eqnarray}
%

%Therefore, one motivates to transform wavefunction as
%
%\begin{eqnarray} \label{eq: psi_transform}
%\Psi(x, t) & \rightarrow & \bar{\Psi}(\bar{x}, t) = \left({ \frac{\hbar}{m} }\right)^{1/4} \Psi(x, t)~,
%\end{eqnarray}
%
%under the above space-coordinate transformation to retain the normalization condition as usual.

%It must be noted that $ \bar{\Psi}(\bar{x}, t) $ too satisfies transformed Schr\"{o}dinger equation (\ref{eq: modifed Sch eq}).
%Noting eq. (\ref{eq: psi_transform}) and the continuity equation
%
%\begin{eqnarray} \label{eq: continuity}
%\frac{\partial \rho(x, t)}{\partial t} + \frac{\partial j(x, t)}{\partial x} &=& 0  ~,
%\end{eqnarray}
%
%one obtains $j(x, t) \rightarrow \bar{j}(\bar{x}, t) = j(x, t)$
%for the transformation of probability current density under the space-coordinate transformation.
In summary we have the following transformations,
\begin{eqnarray} \label{eq: transformations}
x & \rightarrow & \bar{x} = \sqrt{ \frac{m}{\hbar} } x ~,\nonumber \\
t & \rightarrow & t ~,\nonumber \\
p = - i \hbar \nabla & \rightarrow & \bar{p} = - i \bar{\nabla} = \sqrt{ \frac{ 1 }{ m \hbar }}~p  ~,\nonumber \\
\rho(x, t) & \rightarrow & \bar{\rho}(\bar{x}, t) = \rho(x, t) ~,\nonumber \\
j(x, t) & \rightarrow & \bar{j}(\bar{x}, t) = \sqrt{ \frac{m}{\hbar} } j(x, t)~,\nonumber \\
\Pi(x_d, t) & \rightarrow & \bar{\Pi}(\bar{x}_d, t) = \Pi(x_d, t)~,\nonumber \\
v(x, t) & \rightarrow & \bar{v}(\bar{x}, t) = \sqrt{ \frac{m}{\hbar} } v(x, t) ~.
\end{eqnarray}
%
%In addition, every quantity of length unit transforms to $ \sqrt{\hbar/m } $ times that quantity, and
%every quantity of inverse length unit transforms to $ \sqrt{m/\hbar } $ times the quantity.
%
%If one uses the usual normalization condition
%\begin{eqnarray} \label{eq: new normalization}
%\int_{-\infty}^{\infty} d\bar{x} ~ \bar{\rho}(\bar{x}, t) &=& 1~,
%\end{eqnarray}
%in the new coordinate system, then using eq. (\ref{eq: transformations}), one obtains the normalization condition
%\begin{eqnarray} \label{eq: old normalization}
%\int_{-\infty}^{\infty} dx ~ \rho(x, t) &=& \sqrt{ \frac{\hbar}{m} }~,
%\end{eqnarray}
%in the old coordinate system.
%Equation (\ref{eq: old normalization}) shows that probability density is narrower for higher masses.

%== Section ==== Section ==== Section ==== Section ==== Section ==== Section ==== Section ==== Section ==== Section ==== Section

\section{Numerical calculations}

For numerical calculations, transformed parameters of the potential
and wavepacket are taken as $ \bar{a} = 1 $, $ \bar{ \sigma}_0 = 1
$, $ \bar{x}_c = -10 $, $ \bar{k}_0 = 1 $.
%
%********************************************************
%********************************************************
% Figure
\begin{figure}
\centering
\includegraphics[width=10cm,angle=-90]{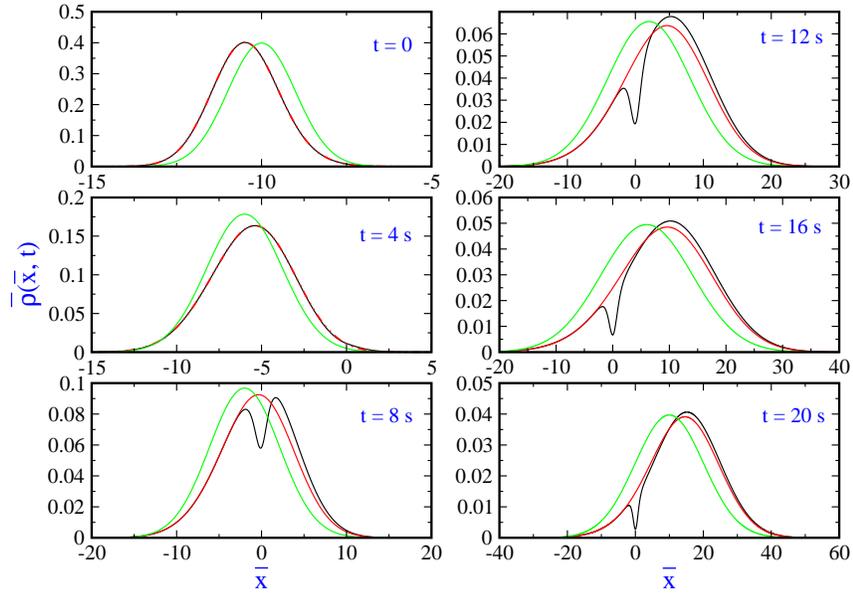}
\caption{(Color online) Transformed probability density
$\bar{\rho}(\bar{x}, t)$ versus transformed space coordinate $\bar{x}$ at six different instants
$ t = 0$, $ t = 4~$s, $ t = 8~$s, $ t = 12~$s, $ t = 16~$s and $ t = 20~$s.
In each plot black curve
shows interacting nonreflecting wavepacket while the red one shows
freely evolving nonreflecting packet and the green one stands for
the propagation of free Gaussian packet.} \label{fig: rho}
\end{figure}
%********************************************************
%********************************************************
In figure \ref{fig: rho}, we have plotted transformed probability
density $\bar{\rho}(\bar{x}, t)$ versus transformed space coordinate
$\bar{x}$ at six different instants.
A gap at the center of the potential is seen
in the plot of interacting nonreflecting wavepacket.
With respect to the relation $ \bar{x}_d = \sqrt{m/\hbar}~x_d $, for
a given width $ \delta \bar{x}$ of $\bar{\rho}$, corresponding width
in $x$ is proportional to the inverse of square root of mass. Thus,
from the forth equation of eq. (\ref{eq: transformations}) it
follows that by increasing the mass, the width of probability
density decreases .
With the above values for the parameters of the wavepacket one
obtains by numerical integration $ N \simeq 2.25 $ for the
normalization constant in eq. (\ref{eq: nonref_packet}), $ \bar{
\langle p \rangle}_{\text{nr}}(0) = \langle \bar{\Psi}_{\text{nr}} |
\bar{p} | \bar{\Psi}_{\text{nr}} \rangle(0) \simeq 1.2222 $ and $
\bar{ \langle x \rangle}_{\text{nr}}(0) = \langle
\bar{\Psi}_{\text{nr}} | \bar{x} | \bar{\Psi}_{\text{nr}} \rangle(0)
\simeq -10.4444 $ for the initial nonreflecting wavepacket. Because
of the relation $ \bar{ \langle p \rangle}_{\text{nr}}(0) > \bar{
\langle p \rangle}_{\text{G}}(0) $, expectation value of space
coordinate of the nonreflecting wavepacket, $ \bar{ \langle x
\rangle}_{\text{nr}} $, in the free-space moves faster than the
center of the Gaussian wavepacket. In addition, in the interacting
case because potential is attractive, wavepacket moves faster when
the wavepacket approaches the well. After, passing the potential it
moves with constat velocity. See Fig.~\ref{fig: ex_del}.
%********************************************************
%********************************************************
% Figure
\begin{figure}
\centering
\includegraphics[width=10cm,angle=-90]{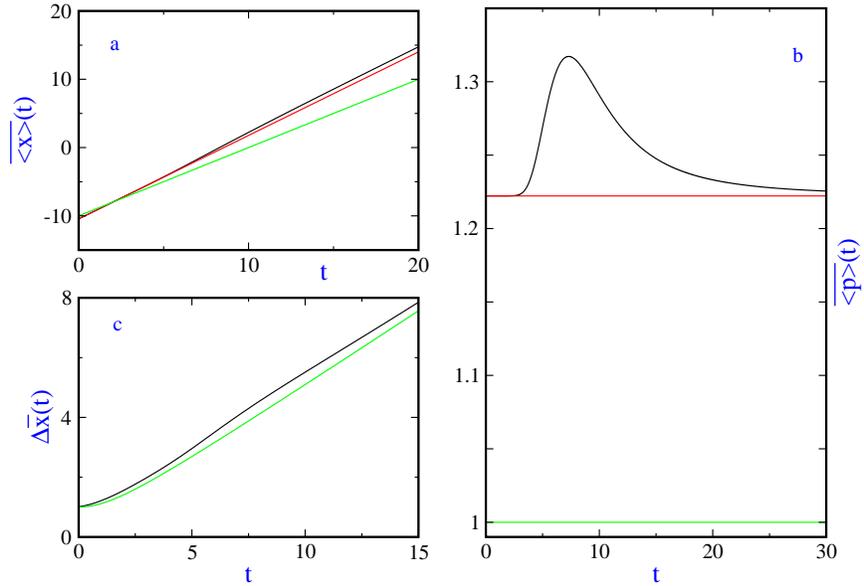}
\caption{(Color online) Expectation value of space coordinate and
momentum (a and b); and uncertainty in position (c)
with the same color scheme as in Fig.~\ref{fig: rho}.}
\label{fig: ex_del}
\end{figure}
%********************************************************
%********************************************************
It must be noted that potential is proportional to the inverse of
mass. Thus, it is more attractive for low masses. So, wavepackets
describing particles with low values of mass move faster in
comparison to the large ones. As one sees in plot b) of figure
\ref{fig: ex_del} uncertainty in position,
$ \Delta \bar{x} = \sqrt{
\langle \bar{\Psi}_{\text{nr}} | \bar{x}^2 | \bar{\Psi}_{\text{nr}} \rangle -
\langle \bar{\Psi}_{\text{nr}} | \bar{x} | \bar{\Psi}_{\text{nr}} \rangle ^2
} $,
is larger for the
interacting nonreflecting wavepacket compared to the freely evolving
Gaussian wavepacket.

\begin{figure}
\centering
\includegraphics[width=10cm,angle=-90]{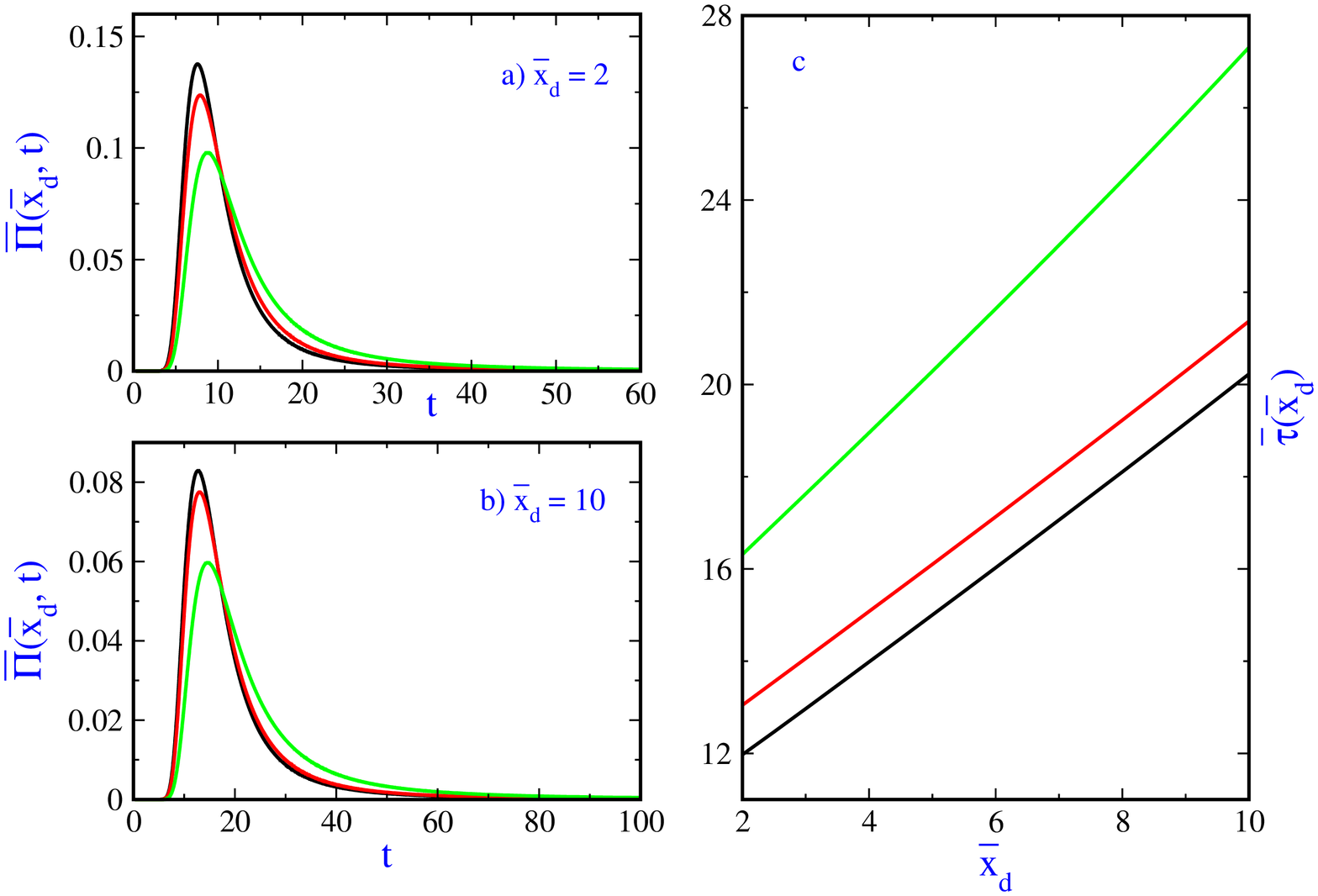}
\caption{(Color online) Arrival time distribution $\bar{\Pi}(\bar{x}_d, t)$
versus time (left column)  and mean arrival time $\bar{\tau}(\bar{x}_d)$ versus
detector position $ \bar{x}_d $
with the same color scheme as in Fig.~\ref{fig: rho}.}
\label{fig: arriv_dis_xd}
\end{figure}

In figure \ref{fig: arriv_dis_xd} we have plotted arrival time
distribution (\ref{eq: arr_dis}) at the detector location $
\bar{x}_d=2 \bar{a} $. Relation $ \bar{x}_d = \sqrt{m/\hbar}~x_d $
means fixing $x_d$ and changing $ \bar{x}_d $ corresponds to
changing mass. From plots (a) and (b) of this figure it is found
that the width-in-time of the arrival time distribution increases
with $ \bar{x}_d $ (mass) in contrast to the width-in-space of the
detection probability density $\rho$ where width-in-space in a given
time decreases with mass. Plot (c) of Fig.~\ref{fig: arriv_dis_xd}
shows the mean arrival time (\ref{eq: mean_arr}). In agreement with
the previous findings, $\bar{\tau}$ is the lowest for the
interacting nonreflecting wavepacket while is the largest for the
free Gaussian. For all types of packets mean arrival time increases
with $\bar{x}_d$ (mass).
%********************************************************
%********************************************************
% Figure
%********************************************************
%********************************************************

\begin{figure}
\centering
\includegraphics[width=10cm,angle=-90]{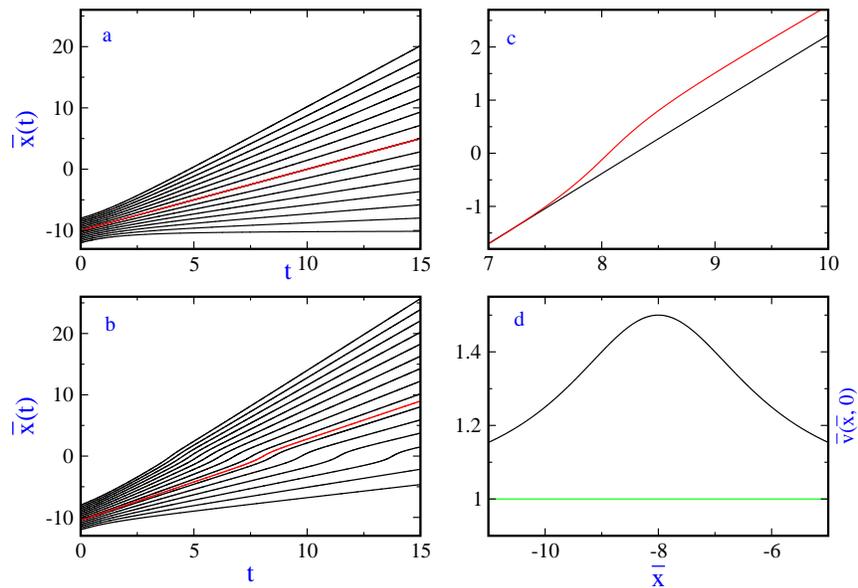}
\caption{(Color online) A selection of Bohmian trajectories
for free Gaussian wavepacket (a) and interacting nonreflecting
wavepacket (b). In both plots (a) and (b) red trajectory starts at
$\bar{\langle x \rangle}(0)$. In plot (c) black curve shows $\bar{\langle x
\rangle}(t)$ for the interacting nonreflecting wavepacket and the red
one represents the Bohm trajectory with initial position $\bar{\langle x
\rangle}(0)$. In plot (d)
one sees initial Bohmian velocity as a function of $\bar{x}$ for
nonreflecting interacting wavepacket (black curve) and for free
Gaussian packet (green curve). }
\label{fig: traj}
\end{figure}

Figure \ref{fig: traj} represents a selection of Bohmian paths. In
plot (a), trajectories (\ref{eq: bohm_tarj_G}) for the free Gaussian
have been plotted, while in plot (c) trajectories are for a particle
which is guided by the wavepacket $\Psi_{\text{i,nr}}(x, t)$. In
agreement with plot (c) of Fig.~\ref{fig: ex_del}, trajectories
spread out more in the case of interacting nonreflecting packet. In
plot (c) we have depicted $ \bar{ \langle x \rangle}(t)$ and the
Bohm path with initial position $ \bar{ \langle x
\rangle}_{\text{nr}}(0)$. It is seen that the Bohm path overtakes
the expectation value of space coordinate.
%and the
%corresponding classical path with initial conditions $ x_{\text{cl}}
%= \bar{ \langle x \rangle}_{\text{nr}}(0)$ and $ v_{\text{cl}} = \frac{
%\langle p \rangle_{\text{nr}}(0) }{ m }$. Since $ \langle V(x)
%\rangle \neq V(\langle x \rangle) $, in agreement with Ehrenfest
%theorem, $ \langle x \rangle(t)$ does not follow classical path.
Plot (d) represents initial Bohmian velocity (\ref{eq:
vb_initial_nonref}) versus space coordinate $\bar{x}$. As one sees,
velocity is higher for the interacting nonreflecting wavepacket
compared to the free Gaussian.

%== Section ==== Section ==== Section ==== Section ==== Section ==== Section ==== Section ==== Section ==== Section ==== Section

\section{Summary and conclusion}

In this work we constructed a nonreflecting wavepacket by
superposing of eigenstates of the reflectionless sech2 potential.
Then, evolution of this wavepacket in the presence of the potential
and in free space studied and compared to the free propagation of
Gaussian wavepacket.
It was found that nonreflecting wavepacket moves faster and its
width, uncertainty in position, is larger.
Mean arrival time at a detector behind the well was
computed as a function of mass. It was seen mean arrival time
increases with mass in agreement with one's intuition and it is the
largest for the free Gaussian compared to the nonreflecting
wavepacket in free space and in the interacting case. Finally, a
selection of Bohmian trajectories were computed.

%********************************************************
%********************************************************
% Figure
%********************************************************
%********************************************************

%********************************************************

%===========================

\end{document}